Original Paper

# A Smartphone App to Support Sedentary Behavior Change by Visualizing Personal Mobility Patterns and Action Planning (SedVis): Development and Pilot Study

Yunlong Wang[1*], PhD; Laura M. König[2*], PhD; Harald Reiterer[1], PhD

[1]Department of Computer and Information Science, University of Konstanz, Konstanz, Germany
[2]Department of Psychology, University of Konstanz, Konstanz, Germany
[*]these authors contributed equally

**Corresponding Author:**
Yunlong Wang, PhD
Department of Computer and Information Science
University of Konstanz
Universitätsstraße 10
Konstanz, 78457
Germany
Phone: 49 7531 88 3704
Email: yunlong.wang@uni-konstanz.de

## Abstract

**Background:** Prolonged sedentary behavior is related to a number of risk factors for chronic diseases. Given the high prevalence of sedentary behavior in daily life, simple yet practical solutions for behavior change are needed to avoid detrimental health effects.

**Objective:** The mobile app SedVis was developed based on the health action process approach. The app provides personal mobility pattern visualization (for both physical activity and sedentary behavior) and action planning for sedentary behavior change. The primary aim of the study is to investigate the effect of mobility pattern visualization on users' action planning for changing their sedentary behavior. The secondary aim is to evaluate user engagement with the visualization and user experience of the app.

**Methods:** A 3-week user study was conducted with 16 participants who had the motivation to reduce their sedentary behavior. Participants were allocated to either an active control group (n=8) or an intervention group (n=8). In the 1-week baseline period, none of the participants had access to the functions in the app. In the following 2-week intervention period, only the intervention group was given access to the visualizations, whereas both groups were asked to make action plans every day and reduce their sedentary behavior. Participants' sedentary behavior was estimated based on the sensor data of their smartphones, and their action plans and interaction with the app were also recorded by the app. Participants' intention to change their sedentary behavior and user experience of the app were assessed using questionnaires.

**Results:** The data were analyzed using both traditional null hypothesis significance testing (NHST) and Bayesian statistics. The results suggested that the visualizations in SedVis had no effect on the participants' action planning according to both the NHST and Bayesian statistics. The intervention involving visualizations and action planning in SedVis had a positive effect on reducing participants' sedentary hours, with weak evidence according to Bayesian statistics (Bayes factor, $BF_{+0}$=1.92; median 0.52; 95% CI 0.04-1.25), whereas no change in sedentary time was more likely in the active control condition ($BF_{+0}$=0.28; median 0.18; 95% CI 0.01-0.64). Furthermore, Bayesian analysis weakly suggested that the more frequently the users checked the app, the more likely they were to reduce their sedentary behavior ($BF_{-0}$=1.49; $r$=−0.50).

**Conclusions:** Using a smartphone app to collect data on users' mobility patterns and provide real-time feedback using visualizations may be a promising method to induce changes in sedentary behavior and may be more effective than action planning alone. Replications with larger samples are needed to confirm these findings.

(*JMIR Form Res 2021;5(1):e15369*)   doi: 10.2196/15369

## Introduction

**Background**

Sedentary behavior refers to any waking behavior characterized by an energy expenditure ≤1.5 metabolic equivalents while in a sitting, reclining, or lying posture [1,2]. Studies have shown evidence of the detrimental effects of prolonged sedentary behavior, which is ubiquitous in daily life, especially when at work. For instance, a study [3] tracking 425 adults for 10 years (2002-2004 to 2012-2014) showed that a greater increase in sedentary behavior was associated with detrimental changes in cardiometabolic risk factors, such as waist circumference, high-density lipoprotein cholesterol, and triglycerides, independent of the change in moderate-to-vigorous physical activity. In other words, exercising after sitting for a prolonged time while at work might not reduce the health risk caused by prolonged sitting. Moreover, a study [4] involving 168 participants in Australia showed that the total number of breaks in sedentary time was associated with improved health parameters, such as significantly lower waist circumference, BMI, triglycerides, and 2-hour plasma glucose. Consequently, several governments (eg, Australia [5] and Canada [6]) have released guidelines to specifically reduce people's sedentary behavior for improved health. For example, people with a sedentary lifestyle could introduce light physical activity (eg, short walking) throughout the day to reduce the risk of many chronic diseases.

The high prevalence of sedentary activities in daily life leads to a stronger habit of sedentary behavior [7], that is, a high degree of automaticity owing to frequent repetition in a stable context [8], which makes it difficult to change in the long term [9]. Interventions are, therefore, needed to support individuals to reduce their sedentary time. In their review, Chu et al [10] divided intervention strategies for reducing sedentary behavior into 3 categories: (1) educational or behavioral (eg, goal setting, action planning, and self-monitoring), (2) environmental changes (eg, sit-stand workstation and treadmill desk), and (3) multi-component (eg, sit-stand workstation plus goal setting). Environmental and multi-component interventions might require policy support and additional facilities, which might hinder their immediate application on a larger scale. Therefore, simple yet practical solutions are needed.

Mobile devices, including smartphones and wearables (eg, smartwatches and fitness wristbands), might be useful platforms for sedentary behavior interventions. First, the prevalence of both smartphone and wearable device ownership is increasing globally [11]. As smartphones include sensors that allow for the collection of physical activity data [12], smartphone owners do not need additional devices to collect data and receive interventions, thus making the solution simple to deliver and practical to use. Second, interest in mobile apps targeting lifestyle behaviors such as physical inactivity is high [13]. Accordingly, research on digital solutions for the promotion of physical activity and the reduction of sedentary behavior is increasing [14]. However, compared with the number of apps targeting physical activity, there are only a few apps specifically targeting sedentary behavior [15]. Moreover, as previous reviews noted, both commercially available apps and apps developed for research are often not grounded in theory, which might limit their effectiveness [16]. This study, therefore, sought to develop a mobile app for sedentary behavior change that is grounded in behavioral theory. Specifically, it sought to integrate the parameters of action planning and the visualization of users' sedentary behavior patterns to better support sedentary behavior change.

**Action Planning for Sedentary Behavior Change**

Wang et al [17] recently proposed a holistic framework for developing digital health behavior interventions, drawing from several classic theories of health behavior change in psychology, such as social-cognitive theory [18,19] and the health action process approach (HAPA; Figure 1) [18]. The latter theory is especially important for the design of health behavior interventions as it bridges the intention-behavior gap through action planning [20]. Several meta-analyses have shown that action planning is positively related to goal attainment and health behavior change [21-23] and thus might be an effective behavior change technique. Accordingly, action planning was included in the taxonomy of behavior change techniques [24]. An action plan combines specific situation parameters (*when* and *where*) and a sequence of actions (*how*) for a target behavior [25]. In this vein, it is suggested that behavior will be triggered automatically when encountering specific situations [26].

Although action planning is an effective behavior change technique, there are only a few studies that included action planning in digital interventions targeting sedentary behavior [27]. In a recent systematic review of digital technologies supporting health behavior change [28], only 2 out of the 45 studies reviewed involved action planning related to sedentary behavior change. On the basis of the step counts at baseline, Aittasalo et al [29] offered participants visual feedback to facilitate action planning, whereas De Cocker et al [30] used several motivational questions to stimulate the participants to make action plans. In both studies, sedentary behavior was successfully reduced. However, both used action planning as one of the several behavior change techniques, and it is, therefore, unclear whether the change can be solely attributed to action planning. Maher and Conroy [7], on the other hand, specifically tested the main effect of action planning on reducing sedentary behavior and found that daily action planning did not induce sedentary behavior change. This study, however, has limitations. First, sedentary behavior was only assessed subjectively, which might not correspond to objectively measured behavior [31]. Second, the quality of the action plans was not evaluated, which might preclude important insights into why the intervention was not successful.

The quality of an action plan can be evaluated based on plan characteristics such as the specificity of the situational parameters; plan instrumentality, that is, the degree to which a plan is helpful to achieve the desired outcome; and viability, that is, how realistic an action plan is. Fleig et al [32] showed that specificity of when to perform a behavior and instrumentality of the action plan were related to an increased likelihood of plan enactment. Quality of action plans, therefore, might be an important variable to consider when evaluating interventions. Although none of the aforementioned studies on sedentary behavior change investigated the quality of action plans, this study aims to test the effect of action planning on sedentary behavior change quantitatively and additionally include a qualitative analysis of the action plans to determine their specificity, instrumentality, and viability.

**Figure 1.** The model of the health action process approach.

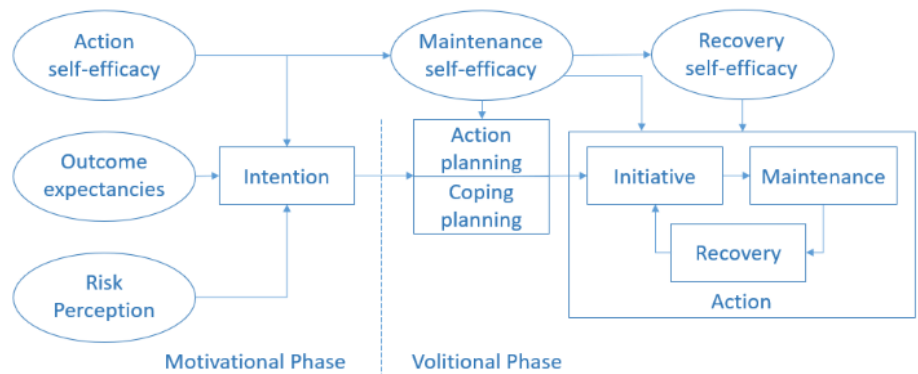

### Visualizations of Mobility Patterns

Human mobility patterns reflect the spatial and temporal periodicity or routines of human activities in their daily lives [33-35]. Mobile devices allow for the passive monitoring of human mobility, including physical activity and sedentary behavior. When fed back to the user, the data might help them to generate meaningful insights about their activity patterns and subsequently induce behavior change. Self-monitoring and feedback based on the collected data are frequently used to change physical activity and sedentary behavior [36-39]. However, the feedback is often numerical or uses simple static visualizations such as bar charts or line graphs to display step counts or energy expenditure (eg, Google Fit and Fitbit). On the basis of this information, it might be difficult to extract all relevant information needed to formulate effective action plans defining the *when*, *where*, and *how*. It could be hypothesized that map-based visualizations, such as visualizations provided by apps to track running, might be more effective, as they provide information about where activities took place [40].

Building upon the idea that visualizations of sedentary behavior data might facilitate action planning [29], a novel tool to support action planning for reducing sedentary behavior using interactive visualization was developed. This study thus extends previous mobile sedentary behavior interventions by using an interactive visualization of sedentary behavior data to specifically support daily action planning, which in turn was hypothesized to reduce sedentary time in daily life. A mobile app, SedVis, was implemented by the study team. Mobility patterns were determined based on objective data collected by the app: using internal sensors of the smartphone and existing services provided by the operating system, SedVis automatically tracks and classifies users' activity (eg, walking, biking, and being in a vehicle), step count, and location. In this vein, it determines locations and time windows in which users are sedentary. The visualization elements thus correspond to the aforementioned action planning factors—when, where, and how (ie, the planned activity). By specifically highlighting situations in which users are sedentary, visualizations can serve as a visual aid for formulating action plans. To the best of our knowledge, SedVis is the first app targeting visualizations and action planning on mobile devices for sedentary behavior change.

### Study Objectives

This paper reports on the results of a 3-week user study of SedVis (N=16). Specifically, the study aims to answer 4 research questions (RQs). The first aim is to examine the effect of SedVis on users' action planning for their sedentary behavior change (RQ1). Specifically, we tested whether using the visualization improved 3 characteristics of action plans that have been identified as potentially impacting the effectiveness of the plans for behavior change [32]: (1) specificity, that is, the level of detail the plan provided on when and where the behavior was to be shown; (2) instrumentality, that is, the degree of helpfulness of the plan for behavior change; and (3) viability, that is, the degree of control an individual has over plan enactment of formulated action plans. Second, we tested whether the intervention involving visualizations and action planning is effective in reducing sedentary behavior compared with action planning without visualizations (RQ2). Third, because the designed visualizations could also serve as a self-monitoring tool, users' engagement with the visualizations in SedVis and its impact on users' sedentary behavior change was investigated (RQ3). Four, user acceptance and experience of SedVis as a simple intervention tool for the daily use of the sedentary population were studied (RQ4).

## Methods

### SedVis App

SedVis was developed and implemented by the study team. Specifically, YW developed the app concept and programmed the app. LK and HR tested the app and provided critical feedback. App development was guided by the following requirements that were derived from the literature: (1) develop an app to reduce sedentary behavior, (2) grounded in HAPA with a focus on action planning, (3) using mobility pattern visualizations, and (4) that is simple and practical to use to ensure user acceptance and use in daily life.

### Data Collection

SedVis was developed for Android smartphones and pretested internally by the study authors. It collected the data of physical activities (via Google Activity Recognition application programming interface [API]) [41], geolocation (via Google Maps API), steps (via Google Fit API), screen states (turned on or off), users' interaction within the app, users' action plans, and time stamps. On the basis of the built-in sensor data, the Google Activity Recognition service on the Android platform could recognize physical activities, including running, walking, cycling, being in a vehicle, and being still. As high battery consumption (eg, through constant geolocation tracking) or large disk-space requirements might lead to users abandoning the app, geolocation was only updated when movements were detected based on activity recognition and steps counting every 5 seconds. In addition, a new data point was only recorded when a change in the activity state was detected (eg, the steps increase or the physical activity changes). This strategy minimizes energy consumption and data storage without losing information on users' mobility [42].

To improve power consumption, Google imposes limitations on background services since Android 8.0. Some original equipment manufacturer versions of Android (eg, Xiaomi MIUI [43] and Huawei EMUI [44]) additionally introduced limitations on background services to optimize the battery life. The operating system might automatically kill the background service. Therefore, the logged data might not indicate the difference between true sedentary periods and the periods during which the background service was not running. Therefore, a timer was added to the background service to log a timestamp to the local database every 20 min. To improve the data collection quality, a data collection service was bound to a notification showing the latest update time, steps, and activity in the notification bar (as shown in Figure 2). A system clock was used to monitor if the background service was running and, if necessary, to initiate a restart. Users could also manually restart the data collection service if the notification disappeared.

**Figure 2.** The always-on notification of SedVis on a user's smartphone.

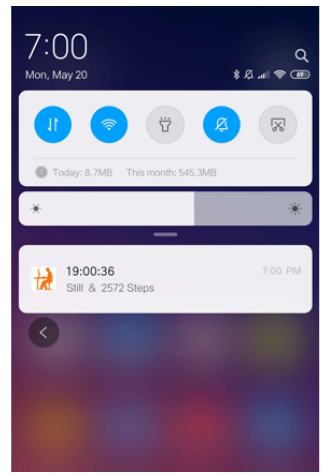

### Mobility Pattern Detection

Mobility patterns refer to when, where, and how the user moves or is sedentary, which directly corresponds to the 3 elements in action planning (ie, when, where, and how). In SedVis, this involved tracking of users' moving trajectory and sedentary place detection. The trajectories showed the routes the user had taken and related information on step counts and time windows. The app detected the users' physical activity every 5 seconds, which enabled a high temporal resolution for trajectory tracking. Modern smartphones use high-precision and low-power movement sensors, which make physical activity recognition and step tracking both accurate and efficient [45]. Google Play services provide fused location tracking by using GPS, Wi-Fi, and cellular signals to allow for precise positioning even in some indoor environments (accuracy depends on the strength of the indoor Wi-Fi and cellular signals).

Custom programmed sedentary place detection was used to detect the participants' sedentary places based on the users' geolocation data. Many office workers spend the day in a limited number of locations (eg, home, office, and lab) where they spend much time sitting. Existing services, such as the Places software development kit for Android [46], only recognize public places (eg, the university), which could not enable personalized place detection in other places such as at home. Therefore, a spatio-temporal data clustering algorithm [47] was used to detect the places based on each user's data. These detected sedentary places, which were displayed in mobility pattern visualizations, provide users with intuitive cues on where to reduce their sedentary behavior.

### Mobility Pattern Visualization

Within SedVis, users could access 2 visualizations of data on their sedentary and active hours that were generated based on the collected mobility pattern data. An hour was labeled sedentary if the user took fewer than 250 steps per hour as in the Fitbit mobile app and according to recent evidence suggesting that 2-min walking (about 250 steps) per hour might lower the risk of premature death [48].

Participants could access the single-day visualization via the dashboard or by clicking the always-on notification (Figure 2). In the daily visualization, the tracked trajectories and the detected sedentary places are shown on a map, and the corresponding temporal information is shown using a bar chart (as shown in Figure 3) for a single day. Specifically, sedentary hours were marked by orange bars, and sedentary locations were marked with orange triangles on a map to highlight situations in which users were sedentary. Participants could interact with the visualization by tapping on the bar chart or on the locations and trajectories displayed on the map. Specifically, they could see (1) the active hours and the corresponding routes on the map once clicking on a blue bar and (2) the sedentary hours and the corresponding locations once clicking on an orange bar. Likewise, clicking the sedentary location on the map highlighted the corresponding sedentary hours in the bar chart. Although the bar chart illustrated temporal patterns, the map demonstrated spatial patterns. Participants could switch between days by tapping on the arrows at the bottom of the screen.

In the multi-day visualization, data were aggregated across multiple study days. Sedentary places were determined based on aggregated data from the user-selected days. Differing from the daily visualization, the bar chart in the multi-day visualization showed the frequency of the user being sedentary in each hour during the selected days for all the places or one selected place (Figure 4).

**Figure 3.** The mobility patterns in the daily visualization mode.

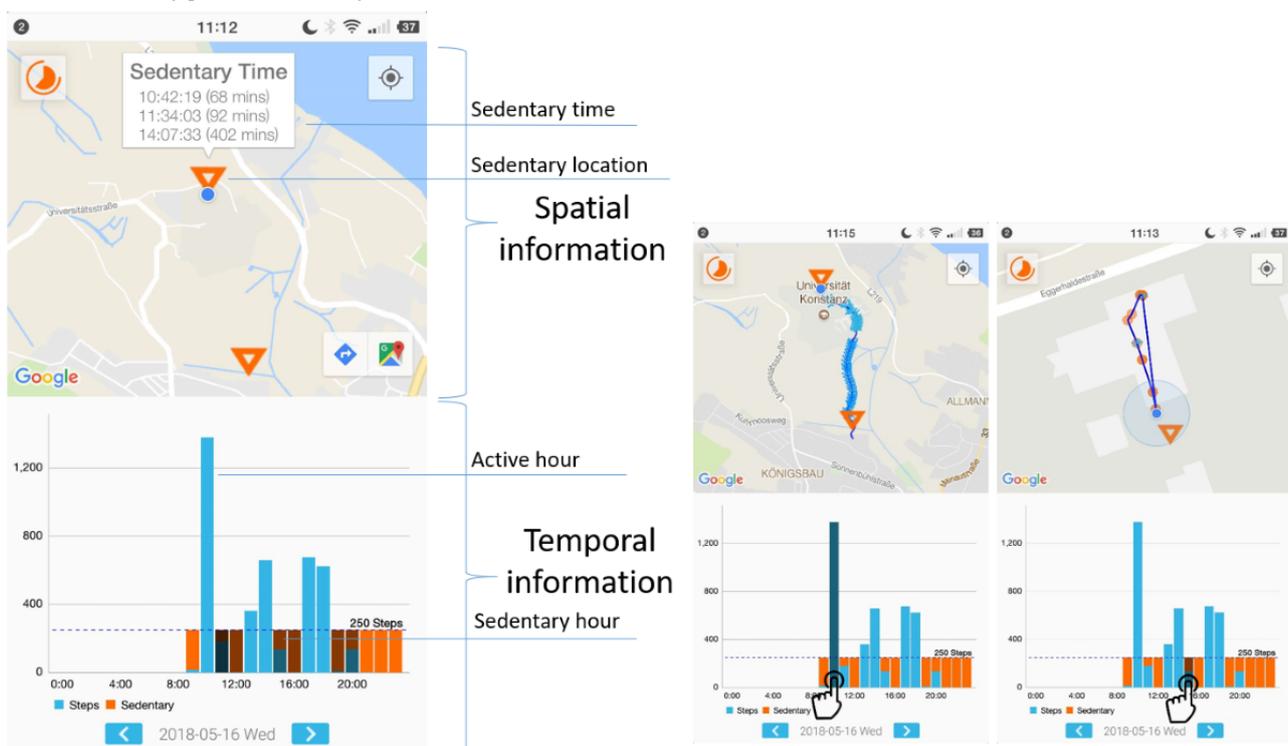

**Figure 4.** The multi-day visualization.

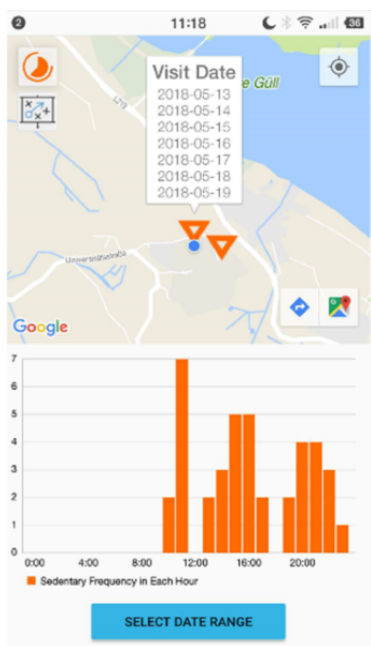

### Action Planning

The user could enter the action planning view from the dashboard, the daily notification, or the shortcuts in the visualization views (see the second button on the left-side corner of Figure 3 and Figure 4). All action plans that the user had made were shown in a list view. The action plans were shown chronologically and could not be deleted. When adding an action plan, the user was asked to specify the *when*, *where*, and *how* elements (Figure 5).

**Figure 5.** The action planning function in the app.

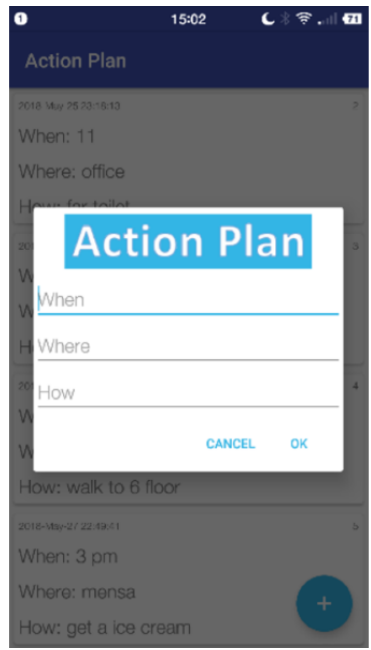

### Dashboard

From the dashboard of SedVis, participants could access all the functionalities of the app, as shown in Figure 6. In the settings tab, the study staff could enable intervention functions. Passwords were used to restrict users' access to these functions during the study.

**Figure 6.** The dashboard of SedVis.

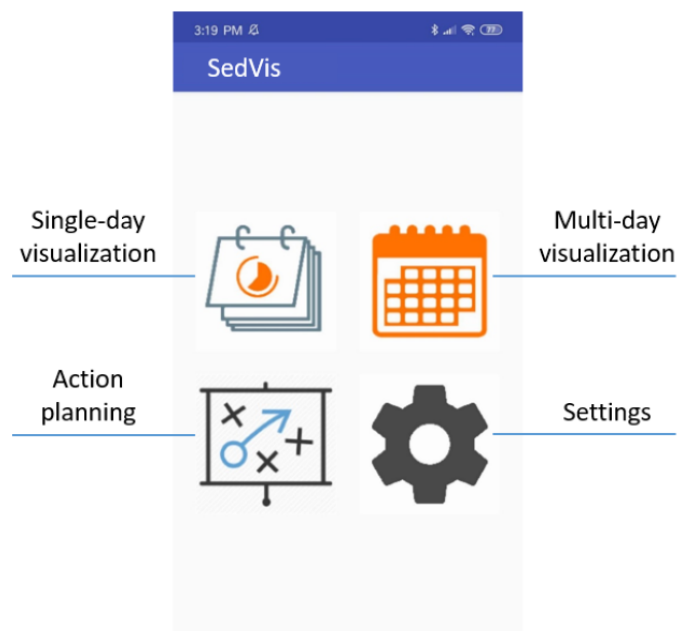

## Study Design and Procedure

The study deployed a mixed design with one between-subject factor *group* (with vs without visualization) and one within-subject factor *time* (baseline vs intervention; Figure 7). Participants were assigned to 1 of the 2 groups, which determined the intervention they received: group A (intervention group), for which the visualization functions were enabled, and group B (active control group), for which the visualization functions were disabled. Participants were assigned to the groups according to the enrollment time (ie, every odd number was assigned to group A, whereas every even number was assigned to group B). This strategy enables fast study deployment for each participant while maintaining the balance of sample size in both groups [49]. As the sequence of the participants enrolled in the study was random, this strategy preserved the randomization of group allocation.

The study included 3 interviews (ie, the entry interview before starting data collection, the after-baseline interview after week 1, and the exit interview after week 3) on day 1, day 9, and day 25 for each participant. The data collected on these 3 days in the app were excluded from the data analysis because they were incomplete and could not be compared between participants because of appointments being scheduled throughout the day.

During the entry interview, the participants were informed about the purpose of the study, signed the consent form, and filled out questionnaires on demographics and psychosocial variables related to sedentary behavior (intention, risk perception, and self-efficacy based on HAPA; only results for intention are reported as it is the only construct directly associated with action planning) [25]. A member of the study staff then installed SedVis on their smartphones.

The after-baseline interview took place on the first day after the baseline week. Participants again filled out the questionnaire on psychosocial variables before watching an educational video about the risks of prolonged sedentary behavior [50]. Subsequently, the study staff showed them a flyer to explain behavior change theory [51] and emphasized the importance of action planning. The participants were asked to make at least one plan per day to reduce their sedentary behavior for the following 2 weeks. Finally, the study staff introduced the functions of the app, depending on which group participants of the session were assigned to. For group A, all the functions were activated, including daily visualization; multi-day visualization, which allowed for displaying mobility patterns for multiple days; and action planning. For group B, only the action planning function was enabled. Participants were demonstrated how to make an action plan in the app with dummy examples (eg, *10 am, office, take a walk*). For both groups, participants were asked to set a daily reminder within the settings of the app when they used it for the first time, which served as a prompt to make action plans.

After 2 more weeks, the participants returned to the lab for the exit interview, when they again completed a questionnaire on psychosocial variables as well as an additional questionnaire on user experience. They further transferred the data stored on their smartphones to the study team by email and took part in a short, semistructured interview. Participants were asked questions about their current health status, especially regarding acute infections that might have limited their physical activity; changes in daily routines that might have affected their physical activity or sedentary behavior; and divergences from their sleeping habits, for example, having slept longer or shorter than usual. In addition, they were provided with a list of their action plans and asked to rate them. The participants' answers were written on printed forms and archived into digital forms after the study. Each participant received €20 (US $25) after completing the study.

The ethics committee of the University of Konstanz approved the study protocol. For privacy reasons, only data related to the study were collected. To ensure transparency of data collection, data were recorded and stored on the participants' smartphones until the study was completed. The participants were shown the data details when they transferred the data via email to the study staff. The data remained anonymous and stored on the encrypted server hosted in the university.

**Figure 7.** The mixed design of the 3-week study.

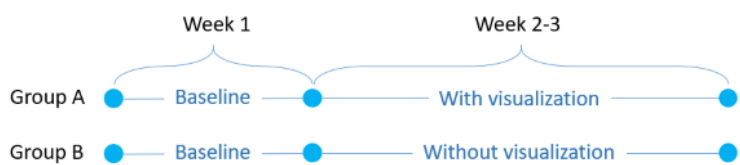

## Participants

Participants were recruited through university mailing lists, the authors' social media profiles, and posters in the university. A total of 16 participants expressed interest in taking part in the study. Participants were eligible for participation if they (1) had the intention to change their sedentary behavior, (2) had no injuries that precluded them from being physically active, (3) were able to speak English fluently, (4) had a smartphone with Android 6.0 and above, (5) did not use a standing desk, and (6) had no travel plans during the study period. The fifth criterion was used to filter out people who had already started to change their sedentary behavior. The other criteria were used to control the motivation and objective ability for using the app, communicating with the study staff, and changing sedentary behavior. The criteria were listed in the study advertisement, and potential participants self-evaluated whether they fit the inclusion criteria. In addition, the intention to change sedentary behavior was assessed in the entry interview as a control measure.

All 16 participants were students (9 out of 16 PhD students and 9 out of 16 females) at the university. Group A comprised 5 females and 3 males. Their mean age was 26.6 years (SD 3.8). Group B comprised 4 females and 4 males. Their mean age was 27.0 years (SD 4.0). Among the 16 participants, one was overweight (ie, BMI>25 kg/m²), one was underweight (ie, BMI<18.5 kg/m²), and the remaining had a normal weight (mean BMI of 22.0 kg/m², SD 2.8).

## Measures

### Sedentary Behavior

Sedentary hours were assessed throughout the 3 weeks of the study and calculated based on the step counts assessed by the SedVis app, which were again determined based on the Google Activity Recognition API native to Android smartphones. Studies have shown that off-the-shelf smartphones and smartwatches could provide a reliable estimation of users' physical activity [45,52]. Sedentary behavior was quantified per hour: an hour was labeled as sedentary if less than 250 steps were recorded.

It should be noted that the sedentary hours the app estimated included the participants' sleeping time. It was assumed that the participants' sleeping time did not change over the 3-week study period, which was confirmed by the participants in the exit interview. Thus, the difference in the daily sedentary hours between the baseline and intervention weeks should not be influenced by the sleeping hours. The estimated sedentary hours will be used to reflect sedentary behavior in the rest of the paper.

*Number of Action Plans*

The total number of action plans formulated during the 2-week intervention phase was counted automatically by the SedVis app. As participants were allowed to repeat the plans of previous days, the number of unique action plans was calculated additionally.

*Quality of Action Plans*

To evaluate the quality of the action plans, the specificity of the when, where, and how of the plans were coded. The rating criteria for 3 levels of specificity (ie, vague, medium specific, and highly specific) were adapted from Fleig et al [32] (see Table 1 for coding criteria). In addition, participants were asked to evaluate the viability (how realistic) and instrumentality (how useful) of their action plans based on the plan characteristics used by Fleig et al [32]. For viability, participants were asked to rate each action plan on a scale from 1 (not realistic at all) to 4 (very realistic). For instrumentality, participants were asked to rate each action plan on a scale from 1 (not helpful at all) to 4 (very helpful).

**Table 1.** Coding criteria for specificity.

| Specificity type | Vague (1[a]) | Medium specific (2[a]) | Highly specific (3[a]) |
| --- | --- | --- | --- |
| "When" | Empty; "Now"; "Anytime"; "Today" | "Every Hour"; "After Lunch" | Timepoint (eg, "13:00") |
| "Where" | Empty; "Out" | Large area (eg, "City," "University") | Places (eg, "Post," "Lab," "Office," "Home," "Library") |
| "How" | Empty | "Going to the park" | Activity (eg, "Walk," "Yoga," "Cycle," "Push-ups," "Stretch," "Stand up") |

[a]The numbers are the rating levels corresponding to vague, medium specific, and highly specific.

*Engagement With the App*

Participants' interaction with the app was quantified by recording all operations in the app during the study, including how often the participants checked the visualizations. In addition, the timestamps of when participants made action plans were logged, which were then used as the basis for discussing the users' experience with the app during the exit interview.

*Intention to Change Sedentary Behavior*

The participants' intention was measured using a scale from 1 ("I do not plan to reduce my sedentary behavior at all") to 4 ("I do exactly plan to reduce my sedentary behavior") following the example in HAPA [25]. The intention was used as a control measure, as participants were required to be motivated to reduce their sedentary behavior instead of other factors (eg, receiving monetary compensation).

*User Experience: Quantitative Measure*

Using the user experience questionnaire (UEQ) [53], the user experience of the app was quantified at the exit interview.

*User Experience: Qualitative Interviews*

In addition, closed- and open-ended questions were used to explore the participants' attitudes toward the app and the study as well as their desired features missing in the app: (1) Would you like to receive a reminder for performing the action plans?, (2) Do you want to continue using this app? Why?, (3) Do you think that the logged data on sedentary time and location were accurate? (only group A), and (4) Did you always take your phone with you during work? Replies were recorded as written notes by the interviewer; most replies were either yes/no or statements of 1 to 2 sentences, for example, "The app underestimated the number of steps because I cannot take my phone with me during experiments." For questions that were usually answered with yes or no, the number of participants replying with either option is reported. Owing to the small sample size and limited number of statements exceeding yes/no, responses were only aggregated if they addressed the exact same issue (eg, the smartphone's sensor not being sensitive enough to properly capture nonsedentary periods); otherwise, individual statements are reported.

*Qualitative Control Measures*

In addition, participants were asked to report unexpected issues that might have affected the data quality during the study: (1) Were your daily routines during the study, including your sleep, typical or not? and (2) Did you have to complete urgent tasks (eg, related to PhD thesis) during the study? This information was recorded to potentially inform interpretation of the data, for example, to explain divergences in step counts between weeks that may mask intervention effects.

**Statistical Analysis**

Data were analyzed using both traditional null hypothesis significance testing (ie, nonparametric Mann-Whitney U tests, Wilcoxon signed-rank tests, and Spearman rank-order correlations to account for the small sample size) and equivalent Bayesian statistics to provide Bayes factors (BFs). In addition, descriptive statistics (ie, median, mean, and SD) were reported, as suggested by Lee et al [54]. RQ1 was evaluated using the Mann-Whitney U test with the independent variable group and dependent variables total and unique number of action plans and measures of action plan quality. RQ2 was evaluated using a Wilcoxon signed-rank test with study conditions as the independent variable and sedentary hours as the dependent variable. RQ3 was evaluated using Spearman rank-order correlation test to examine the relationship between the frequency of checking the visualization and the sedentary hours. RQ4 was evaluated using the Wilcoxon test with the independent variable group and the UEQ scores as dependent variables. JMP Pro (version 14.1.0 [55]) was used for statistical analysis. The normalized statistics (ie, the $Z$ scores) of the nonparametric null-hypothesis significance tests will be reported along with the $P$ values.

We adopted BFs as complementary statistics. The conventional null hypothesis significance tests provide little information when the result is not statistically significant; only the alternative hypothesis is tested [56]. Nonsignificant results might support a null hypothesis over the alternative, or the data are just insensitive. In contrast, BFs [57] compare the extent to which the samples support 2 hypotheses (eg, equal or different). Moreover, Bayesian methods allow more principled conclusions from studies with a small sample size of novel techniques in the field of human-computer interaction [58]. Therefore, BF was used in addition to the $P$ value [59] and Cohen $d$ [60] to report and interpret the results. JASP (Jeffreys' Amazing Statistics Program; version 0.9.2) [61] was used for Bayesian analysis.

BF is the ratio of likelihood probabilities. $P(data \mid H_0)$ is the probability of the null hypothesis ($H_0$) given the data, whereas $P(data \mid H_1)$ is the probability of the alternative hypothesis ($H_1$) given the data. The definition of BF is shown in formula 1 below:

$$BF_{01} = \frac{P(data \mid H_0)}{P(data \mid H_1)} = \frac{P(H_0 \mid data)}{P(H_1 \mid data)} \times \frac{P(H_1)}{P(H_0)} \text{ or } BF_{10} = \frac{1}{BF_{01}}$$

BF indicates which hypothesis is supported more by the data. Figure 8 shows the BF classification and the adapted interpretation [62]. The default prior distributions of the alternative hypothesis and the calculation methods for different study designs can be found in the study by Rouder et al [63,64].

The default Cauchy distribution, $r = 1/\sqrt{2}$ was used as the prior distribution when estimating the effect size. Following the JASP guidelines [62], the posterior median and the 95% CI of the effect size are also reported. For correlation analysis, the Bayesian Pearson correlation test was used with the default prior distribution suggested by Rouder and Morey [65]. Depending on the context, the one-side BF ($BF_{-0}$ or $BF_{+0}$) or the two-side BF ($BF_{01}$) is reported.

**Figure 8.** A graphical representation of a Bayes factor classification and interpretation. BF: Bayes factor; H0: null hypothesis; H1: alternative hypothesis.

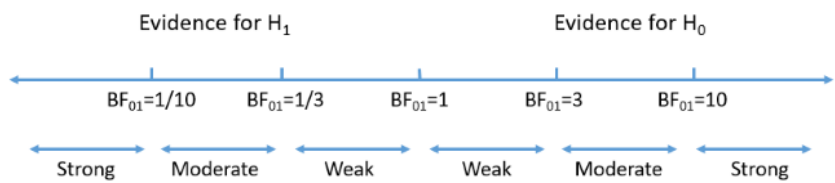

## Results

### Data Collection

All participants completed the study. First, data quality was checked based on the actual running duration of the app to ensure that all participants had access to the app as expected. The missing duration may be caused by the smartphone being switched off or the background service being shut down for battery optimization. Only the data from one participant in group A (A8) showed a relatively low coverage (65.61% of the study duration); for the other participants, the mean coverage was 93.88% (SD 5.44%). After checking the data of the participant A8, it was found that they had the habit of shutting down the phone during the night. Therefore, the missing data did not limit conclusions about the mobility of the participant, and the participant's data were analyzed as planned. No participant reported urgent tasks or long-term travel that might have impacted their daily routines during the study. Therefore, it is unlikely that the potential sedentary behavior change during the study was owing to reasons unrelated to the intervention.

### Participants' Intention (Control Measure)

The participants' intention to reduce sedentary behavior was generally high (median 3.00, mean 3.20, SD 0.59) in both groups. No significant difference was found between groups at each appointment according to Mann-Whitney U tests (appointment 1: $Z=0.06$, $P=.95$; appointment 2: $Z=0.35$, $P=.72$; appointment 3: $Z=0.06$, $P=.95$). The BFs showed evidence preferring $H_0$ (appointment 1: $BF_{01}=2.34$; appointment 2: $BF_{01}=2.23$; appointment 3: $BF_{01}=2.34$). The results indicated that the participants in both groups had similar strong intentions.

### RQ1: Effect of Visualization on Participants' Action Planning

The first aim was to investigate the effect of visualizations on participants' action planning. Both the quantity and quality of the action plans were evaluated.

The Mann-Whitney U test showed no significant group difference regarding the total number of action plans made in the 2 groups ($Z=-0.37$; $P=.71$; median$_{group\ A}$ 8.5; mean$_{group\ A}$ 8.9, SD$_{group\ A}$ 5.69; median$_{group\ B}$ 5.5; mean$_{group\ B}$ 7.8, SD$_{group\ B}$ 6.76). BF ($BF_{01}=2.24$; median 0.11; 95% CI –0.66 to 0.92) showed weak evidence toward no difference ($H_0$). This was the same case for the number of unique action plans ($Z=-1.06$; $P=.29$; $BF_{01}=1.81$; median 0.28; 95% CI −0.50 to 1.14; median$_{\text{group A}}$ 5.0; mean$_{\text{group A}}$ 3.8, SD$_{\text{group A}}$ 2.19; median$_{\text{group B}}$ 2.5; mean$_{\text{group B}}$ 2.8, SD$_{\text{group B}}$ 2.38).

The quality of the action plans showed mixed results, as shown in Table 2. The means of the perceived viability and instrumentality were slightly higher in group B than in group A. The results of the Mann-Whitney U test suggested a statistically significant difference in perceived viability between group A and B. BF also showed weak evidence toward difference ($H_1$) for perceived viability, whereas it suggested no difference ($H_0$) for perceived instrumentality. In addition, no meaningful group differences were found regarding specificity (*When* and *Where*). The means of the specificity of the response activity (*How*) were both very high in the 2 groups because most of the users simply specified the activity as walking.

Table 2. The measurements of the quality of the action plans.

| Measure | Group A, mean (SD) | Group B, mean (SD) | $Z^a$ | $P$ value[b] | $BF^d_{01}$ | Median | 95% CI |
|---|---|---|---|---|---|---|---|
| Perceived viability | 3.28 (0.68) | 3.81 (0.37) | 1.98 | *.048*[c] | 0.71 | −0.67 | −1.72 to 0.17 |
| Perceived instrumentality | 3.10 (0.55) | 3.22 (0.73) | 0.53 | .60 | 2.13 | −0.15 | −1.00 to 0.59 |
| Specificity (when) | 2.55 (0.70) | 1.88 (1.00) | −1.11 | .27 | 1.05 | 0.51 | −0.30 to 1.50 |
| Specificity (where) | 2.21 (0.82) | 2.10 (0.91) | −0.11 | .91 | 2.24 | 0.10 | −0.68 to 0.94 |
| Specificity (how) | 2.99 (0.04) | 3.00 (0.00) | 0.88 | .38 | —[e] | — | — |

[a]Z refers to the normalized statistic of Mann-Whitney U test.

[b]$P$ value of Mann-Whitney U test.

[c]An italicized $P$ value indicates significant difference ($P<.05$).

[d]BF: Bayes factor.

[e]—: For specificity (*how*), no results are reported for BF because the SD in group B was 0.

Therefore, regarding RQ1, no statistically significant effects of the visualizations in SedVis on the participants' action planning were found, except for perceived viability. BFs indicated weak evidence toward no difference between the 2 groups, except for perceived viability.

Regarding the specificity (*When*) of action plans, some unexpected patterns were observed, especially in group B. Two participants (B3 and B6) in group B always entered the current time when they made the plan. They explained in the exit interview that each of their plans was actually what they were about to do at the moment when they logged the plan. Participant B6 further commented that she found it difficult to make action plans for the future because she was not sure about her behavior patterns. In addition, another 3 participants always used vague cues to specify the *When*: participant B1 used *today*, participant B4 used *anytime*, and participant A5 used *today*. Table 3 shows a summary of the *When, Where,* and *How* in the participants' action plans.

Table 3. Summary of "When," "Where," and "How" components identified in the participants' action plans.

| When | Where | How |
|---|---|---|
| • Timepoint (eg, 4 AM)<br>• Now<br>• Vague time (eg, today, tomorrow, and anytime) | • Workplace (eg, university, lab, library, campus, garden, office, building Z, and outdoor)<br>• Home/dormitory/kitchen<br>• City<br>• Park | • Walk (eg, tea walk, walk to post, walk between lectures, walk after lecture/meeting/lunch/dinner, 5-min walk, 6000 steps, and 250 steps per hour)<br>• Yoga<br>• Cycle instead of the bus<br>• Push-ups<br>• Get up and stretch<br>• Stand up for 5 min every 30 min<br>• Jump |

### RQ2: Changes in Participants' Sedentary Behavior

Wilcoxon signed-rank tests showed a marginally significant decrease in daily sedentary time in group A ($Z=-11.50$; $P=.06$) and no significant difference in daily sedentary time in group B ($Z=2.50$; $P=.59$). The median change in sedentary time was −0.19 hours per day in group A and 0.07 hours per day in group B. The results of the Bayesian paired samples $t$ test suggested (with weak evidence) that the daily sedentary hours decreased from the baseline week to the intervention weeks in group A ($BF_{+0}=1.92$; median 0.52; 95% CI 0.04-1.25). This is also mirrored in the descriptive statistics plotted in Figure 9 (mean −0.40, SD 0.63). In contrast, in group B, it was more likely that the intervention had no effect than a positive effect with moderate evidence ($BF_{+0}=0.28$; median 0.18; 95% CI 0.01-0.64; mean 0.17, SD 1.65).

Therefore, regarding RQ2, the intervention involving visualizations and action planning in SedVis had a positive effect on reducing participants' sedentary hours, with weak evidence. Meanwhile, action planning alone had no effect on reducing participants' sedentary hours, with moderate evidence.

**Figure 9.** The participants' daily sedentary hours based on the app-logged data during the baseline week and the intervention weeks. The bars refer to the confidence intervals with 95% confidence level.

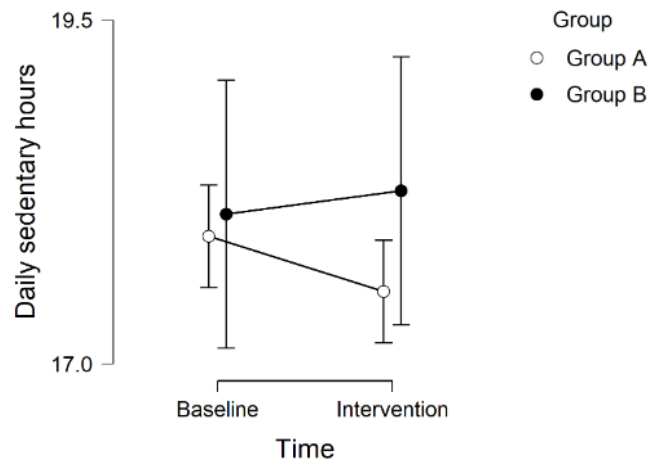

### RQ3: Participants' Interaction With SedVis

The frequency of checking the visualizations per day reflects the participants' strength of self-monitoring, which might also act as a cue for self-reminding of sedentary behavior change. Figure 10 shows the daily frequency of participants checking the visualizations in SedVis. Participants were more likely to check the visualizations from the notification bar (302/442, 68.3%) than from the dashboard (140/442, 31.7%).

To test the assumption that the participants' engagement with the app is positively associated with the effect of the app on their behavior, the daily frequency of participants checking the visualization in SedVis was correlated with their change of sedentary hours, calculated as daily sedentary hours during the intervention weeks minus the counterparts during the baseline week (Figure 11). A Spearman rank-order correlation test did not show a statistically significant correlation between participants' checking of visualizations in SedVis with the change in daily sedentary hours ($\rho=-0.37$; $P=.15$) [66]. Then, a Bayesian Pearson correlation with the alternative hypothesis of negative correlation was calculated. BF (BF =1.49; $r=-0.50$) weakly suggested that the 2 factors were more likely to be negatively related than unrelated. To some extent, this confirmed that the participants' engagement was positively related to the effect of reducing sedentary hours.

**Figure 10.** The daily frequency of participants checking the visualizations in SedVis through the notification and the home screen.

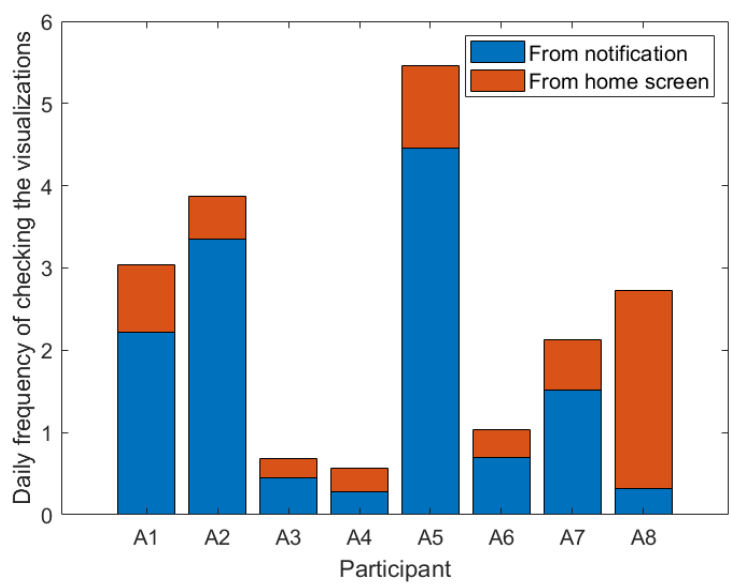

**Figure 11.** The scatter plot of the participants' daily frequency of checking the visualizations and their change in daily sedentary hours in group A. The change of daily sedentary hours (x-axis) equals to the daily sedentary hours during the intervention weeks minus the counterparts during the baseline week. Thus, negative values indicate a reduction in sedentary behavior.

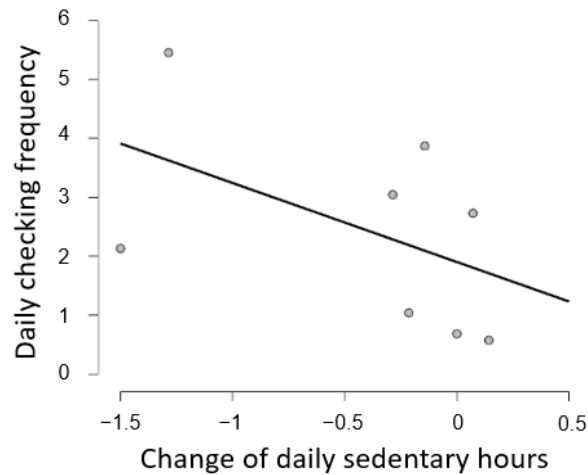

### RQ4: User Experience

#### Quantitative Results

User experience was investigated both quantitatively and qualitatively. By comparing the ratings with the benchmark provided by the UEQ toolkit [10], the participants' scores of user experience were mapped to quality levels, as shown in Table 4.

According to the results of the Bayesian *t* test with the alternative hypothesis that group A scored higher than group B, visualizations yielded more perceived stimulation (BF =1.99; median 0.62; 95% CI 0.05-1.62) and novelty (BF =7.44; median 1.03; 95% CI 0.16-2.16). For other aspects, the scores tended to be equivalent. It was observed that the previewed dependability is only *above average*, which indicates that the participants did not think the data shown in the app were very accurate.

**Table 4.** The user experience scores based on user experience questionnaire.

| User experience questionnaire scales | Group A, mean (SD) | Comparison with benchmark (group A) | Group B, mean (SD) | Comparison with benchmark (group B) | $Z^a$ | $P$ value[b] |
|---|---|---|---|---|---|---|
| Attractiveness | 1.65 (0.69) | Good | 1.44 (0.71) | Above average | –0.74 | .46 |
| Perspicuity | 2.25 (0.81) | Excellent | 2.10 (0.46) | Excellent | –0.95 | .34 |
| Efficiency | 1.91 (0.48) | Excellent | 1.84 (0.79) | Excellent | 0.00 | >.99 |
| Dependability | 1.34 (0.55) | Above average | 1.22 (0.95) | Above average | –0.53 | .60 |
| Stimulation | 1.66 (0.53) | Good | 1.16 (0.63) | Above average | –1.56 | .12 |
| Novelty | 1.22 (0.65) | Good | 0.44 (0.48) | Below average | –2.28 | *.02* [c] |

[a]$Z$ refers to the normalized statistic of Mann-Whitney U test.

[b]$P$ value of Mann-Whitney U test.

[c]An italicized $P$ value indicates significant difference ($P<.05$).

#### Qualitative Results

Although participants were asked to make at least one action plan every day during the 2-week intervention phase, the average number of daily action plans was only 0.59, which hints that participants might not have used the app regularly. According to the feedback in the exit interview, no participant complained about interruptions of daily activities through using the app, although one participant commented that making action plans every day was boring.

The topics of continued use of the app and reminders were often related to participants' responses. Of the 16 participants, 8 wanted to continue using the app to reduce sedentary behavior. The reported reasons for continuing to use the app included "self-monitoring is helpful/could increase self-awareness" (n=4), "I wanted to frequently check step counts" (n=2), "I wanted to see the change" (n=1), and "writing down the action plans are important" (n=1). On the other hand, 4 participants said they did not want to continue to use the app for the following reasons: (1) "it underestimates my steps," (2) "I do not want to always keep the GPS on," (3) "the app provided too little new information compared to other devices like a smartwatch," (4) "I need a reminder for enacting my plans." The remaining 4 participants were undecided about future use. Two participants stated that they would continue to use the app if reminders were implemented; one participant desired a greater range of functionalities, and one participant would have considered continued use if the app would consume less battery and would not require daily action planning. In contrast to the participants who suggested implementing reminders, 6 out of the 8 participants who wanted to keep using the app reported that they did not need a reminder. One of these participants explicitly gave the reason that they did not want to be interrupted during work.

In addition, the topics of accuracy and constantly carrying the phone were linked. Among the 8 participants in group A who had access to the visualization in the app, 4 reported that the data shown in the app seemed accurate. Two participants felt that the app underestimated their steps. This may have been because they did not carry their phone at all times (eg, working in their lab) or because they thought that the sensor in their phone was not sensitive enough. Interestingly, one participant reported that the underestimated steps did not influence their perception of accuracy as they knew the reason, whereas another participant thought that the underestimated steps were disappointing. Therefore, future versions of the app should consider integrating more data sources (eg, wristband or manual adjustment) to improve the users' perceived truthfulness.

## Discussion

### Principal Findings

This paper presents a pilot test of SedVis, an app-based sedentary behavior intervention that aims to reduce sedentary behavior through a combination of mobility pattern visualization and daily action planning. Specifically, it was hypothesized that mobility pattern visualization would lead to improved action plans, which would, in turn, lead to a reduction in sedentary hours.

Contrary to this expectation, the visualizations did not impact the participants' action planning (see the *Results* section). However, these results are in line with those of Maher and Conroy [7], who also found no effect of daily action planning on reducing sedentary behavior in the short term among college students. Furthermore, the data suggested that sedentary behavior change did not correlate with the quantity and quality action plans. As explained by Maher and Conroy [7], one reason for the ineffectiveness could be that the cue-to-action response expected by action planning relies much on the conscious self-regulatory process, which is difficult for highly habitual behavior, such as sedentary behavior. Another explanation could be based on prospective memory [67], inspired by the work of Grundgeiger et al [68]: prospective memory tasks, which require us to remember to do something at a future time, are very difficult, especially when focusing on other tasks. As sedentary behavior is often coupled with other tasks demanding attention, the action plans for reducing sedentary behavior might be easily forgotten.

Still, SedVis may be effective in reducing sedentary behavior: when having access to mobility pattern visualizations, the intervention group slightly reduced sedentary hours compared with baseline. At the same time, the control group did not show a reduction in sedentary hours. It could thus be concluded that visualizations might have impacted sedentary behavior by promoting awareness when self-monitoring sedentary behavior

[69]. This idea is supported by the association between the change in sedentary time and the participants' engagement with SedVis. Engagement with the app, in turn, might have been strengthened by the stimulation and novelty of the visualizations. As Perski et al [70] pointed out in their review on engagement with behavior change interventions, novelty is positively related to engagement as it prevents boredom. The inclusion of novel and stimulating visualizations may thus indirectly influence behavior change.

The participants' evaluations of SedVis with visualizations were good or excellent regarding attractiveness, perspicuity, efficiency, stimulation, and novelty. Only perceived dependability was above average. This may reflect some participants' concerns that SedVis underestimated their steps. At the exit interview, several participants mentioned that they believed the app missed part of their daily steps because they did not take the smartphone with them for certain activities (eg, working in the lab). This limitation of this study could be avoided in future studies by using wearable sensors (eg, wristbands and posture monitors) [71].

The results of this study support the notion that smartphone apps might be an effective tool to reduce sedentary behavior in daily life [15,72]. However, they also indicate that behavior change techniques might differ in their effectiveness to induce changes in sedentary behavior. Three commonly used behavior change techniques were used in this study, that is: self-monitoring, feedback, and action planning [36,72]. Interestingly, action planning was not sufficient to induce changes in sedentary hours in the active control group, whereas additional feedback visualizations induced a small reduction in sedentary hours in the intervention group. Thus, it could be concluded that engaging visualizations to provide feedback on behavior might be more effective in inducing a change in sedentary behavior than action planning. However, as the sample of this study was small, further studies are needed to identify which behavior change techniques are most effective for sedentary behavior change.

### Implications for Future Work

#### Rethinking Action Plans

Although most participants made action plans in accordance with the format of specifying *When*, *Where*, and *How* to reduce their sedentary time, one participant additionally enclosed other contextual cues in their plans, for example, "15:00, lab, take a walk in between experiments" and "13:00, university, walk between lectures." Due to the additional cues—experiment and lectures—the plans might be easier to remember. These plans are in line with the *if-then* format of implementation intentions, which emphasize the contextual cues linking to the goal-directed behavioral response [26,73]. As sedentary behavior is prevalent, the cues of *When* and *Where* might provide limited strength of conditional links to the response behavior. Owing to the requirement of less self-regulatory resources, the more contextual plans in the *if-then* format might be more effective than the plans in *when, where, and how* format [7,73]. However, no prior studies have assessed potential differential effects in sedentary behavior change.

Relating to SedVis, future work might explore how the app could support personalized implementation intentions and their effectiveness on sedentary behavior change, such as generating recommendations of plans based on users' mobility patterns and context, which they might not even notice. Several heuristic rules can be used, for example, going to the restroom downstairs instead of the nearest one or more frequently going to the kitchen to drink water. Armitage [74] found that experimenter-provided and self-generated implementation intentions could be equally effective in reducing alcohol consumption. It is worth investigating this effect on sedentary behavior change following the study design. Some participants commented that making plans every day was boring, so generating plan recommendations might also increase user acceptance in the long term.

*Rethinking Self-Monitoring, Feedback, and Reminders*

As this study suggests that a higher interaction frequency could lead to a greater reduction of sedentary behavior, future work might need to study more convenient and intuitive user interfaces (eg, glanceable feedback [75]) to simplify self-monitoring and interaction with the app even further. In the current version of SedVis, the easiest way to access the daily visualization was to swipe down the notification bar and click on the notification. In a future version, the app could display real-time sedentary information using an always-on progress bar [76] embedded in the notification or the app widget on the smartphone's home screen.

Future work should also consider the users' need for reminders. Participants expressed differential attitudes toward reminders: some of them expressed that reminders for the action plans they made would be helpful because they sometimes forgot the plans; others thought that reminders would be unnecessary because of the potential interruption. Although fixed-time reminders (eg, prompts on PC screens) were frequently used in prior interventions to reduce sedentary behavior at work [77], no studies have explored the effectiveness and user experience of reminders for personalized action plans.

## Limitations

This study has several limitations. First, this study determined sedentary hours based on activity tracked with the smartphone, which may be less rigorous than using dedicated activity trackers (eg, activPAL and ActiGraph) [45]. However, having to wear additional devices might be inconvenient for users (eg, charging the device and attaching the device to the thigh) and may create bias. Moreover, the sedentary hours based on the app-logged data might underestimate the participants' movements. One reason for this might be that participants might not take the smartphone with them during certain activities, such as going to the washroom. Another reason could be that some activities could not be recognized and counted as steps. For example, one participant made an action plan to perform push-ups at home, which cannot be recognized and recorded using a smartphone's sensors. We consider integrating more data sources (eg, smartwatches) in the following version of the app.

Second, the app did not differentiate sleeping time from the sedentary time, and it was assumed that the participants' sleeping time was consistent during the study. Although participants were asked if their sleeping time was normal in the exit interview, their recall might not be accurate. Moreover, the app was not able to distinguish between prolonged periods of sitting and standing. Although using a standing desk at work was an exclusion criterion for participation, it cannot be excluded that participants stood for longer periods, for example, while cooking at home. Future studies, therefore, need to employ more accurate measures for body position to distinguish sitting from lying down and standing (eg, using several sensors [78]).

Third, the small sample size and the relatively large between-subjects variances of the measurements may have reduced the statistical power of the null hypothesis significance tests and may hinder the generalization of our study results. Finally, the study period is relatively short, which limits the validation of the results in short-term scenarios. Therefore, future studies should replicate the results of this study in larger samples and with longer study durations.

## Conclusions

This paper presents the results of a user study in which the effect of a novel visualization within a mobile app on users' action planning and sedentary behavior change was evaluated. The results suggest that using a smartphone app to collect mobility data and provide real-time feedback using visualizations is a promising method to induce changes in sedentary behavior and may be more effective than action planning alone. Future research should thus further explore the potential of the visualizations of users' sedentary behavior to induce behavior change.

XSL•FO
RenderX

## Abbreviations

**API:** application programming interface
**BF:** Bayes factor
**HAPA:** health action process approach
**NHST:** null hypothesis significance testing
**RQ:** research question
**UEQ:** user experience questionnaire

*Edited by G Eysenbach; submitted 04.07.19; peer-reviewed by A Stephenson, J Mair, X Ren, X Ding; comments to author 02.12.19; revised version received 22.04.20; accepted 18.11.20; published 18.01.21*

*Please cite as:*
*Wang Y, König LM, Reiterer H*
*A Smartphone App to Support Sedentary Behavior Change by Visualizing Personal Mobility Patterns and Action Planning (SedVis): Development and Pilot Study*
*JMIR Form Res 2021;5(1):e15369*
*URL: http://formative.jmir.org/2021/1/e15369/*
*doi: 10.2196/15369*
*PMID:*